\documentclass[a4paper,11pt]{article}
\usepackage{pos}

\title{Approximating Feynman integrals using complete monotonicity and Stieltjes properties}

\author*[a,b]{Sara Ditsch}
\author[a]{Johannes M. Henn}
\author[a]{Prashanth Raman}

\affiliation[a]{Max-Planck-Institut für Physik, Werner-Heisenberg-Institut,\\ Boltzmannstr. 8, 85748 Garching, Germany}

\affiliation[b]{Technical University of Munich, School of Natural Sciences, Physics Department, \\ James-Franck-Straße 1, 85748 Garching, Germany}

\emailAdd{ditsch@mpp.mpg.de}
\emailAdd{henn@mpp.mpg.de}
\emailAdd{praman@mpp.mpg.de}

\abstract{We present two novel approaches for the numerical evaluation of Feynman integrals based on their universal analytic properties related to positivity, namely complete monotonicity (CM) and Stieltjes properties. Building on recent results, we exploit the fact that scalar Feynman integrals in the Euclidean region are completely monotonic functions, meaning that all their derivatives have a fixed sign. Building on this observation, the CM bootstrap allows one to reconstruct integrals from differential equations without explicit boundary data, yielding rigorous bounds.
The second method is based on a refinement of CM. We prove that Feynman integrals, within a certain range of parameters, are not only CM but in fact Stieltjes functions. This enables the use of Padé approximants with provable convergence properties in the cut complex plane, providing an efficient method for analytic continuation and fast numerical evaluation. We illustrate the method with simple examples such as the massive bubble integral and discuss applications to multi-loop integrals, including the 20-loop banana integral. Finally, we comment on a number of extensions of these novel avenues for computing Feynman integrals. This talk is based on \cite{Ditsch:2025rdx}.}

\FullConference{Loops and Legs in Quantum Field Theory (LL2026)\\
12-17, April, 2026\\
Bayreuth, Germany\\}

\begin{document}
\maketitle

\section{Introduction}

The numerical evaluation of multi-loop Feynman integrals is a central problem in perturbative quantum field theory. A variety of methods have been developed, including sector decomposition, differential equation techniques, auxiliary mass flow, and Monte Carlo integration, see for example
\cite{Binoth:2000ps,Heinrich:2023til,Smirnov:2021rhf,Borinsky:2023jdv,Hidding:2020ytt,Armadillo:2022ugh,Prisco:2025wqs,Mezzarobba2025Dynaverse,Liu:2022chg}. Despite these advances, efficient and robust evaluation of multi-loop integrals across kinematic regions remains challenging.

In this contribution, we present two novel methods for approximating Feynman integrals based on universal analytic properties related to positivity. The first is based on the observation that scalar Feynman integrals in the Euclidean region satisfy \textit{complete monotonicity} (CM), which was shown in \cite{Henn:2024qwe}. Combining this with the differential equations satisfied by a set of integrals, yields powerful constraints on the space of possible solutions and can uniquely determine the integrals. We extend this idea, by proving that, under suitable conditions, Feynman integrals belong to the class of \textit{Stieltjes functions}, which provides an even stronger constraint. These properties can be leveraged to obtain Padé approximants with rigorous error bounds.

\section{Bootstrapping Feynman integrals from complete monotonicity}
\subsection{Complete monotonicity and Feynman integrals}
A  function $f: R \to \mathbb{R}$ is called \textit{completely monotonic} in the region $R$ if it satisfies (cf. ref.~\cite{widder1941laplace})
\begin{equation}    \label{eq:CM}
    (-1)^n \frac{d^n}{dx^n} f(x) \geq 0\,, \quad \forall \, n \in \mathbb{N}_0, \; \forall \, x \in R \,.
\end{equation}
It has recently been shown in \cite{Henn:2024qwe} that scalar Feynman integrals in the Euclidean region are completely monotonic functions of appropriate kinematic variables. 
This property encodes non-trivial positivity constraints that can be used to constrain the function.

\subsection{The CM bootstrap}

The CM bootstrap combines differential equations for Feynman integrals with positivity constraints from complete monotonicity. Feynman integrals satisfy systems of differential equations of the form
\begin{equation}
\frac{d}{d x} {\bf f}(x) = A(x) {\bf f}(x) \,,
\end{equation}
where ${\bf f}(x)$ denotes a vector of master integrals. Traditionally, solving such systems requires boundary conditions. In the CM bootstrap approach, boundary conditions are replaced by positivity constraints. The differential equation allows rewriting the CM property as a matrix equation for scalar Feynman integrals of the form
\begin{align}
\label{eq:constraint}
    Q_n(x){\bf f}(x)\geq 0\,,\quad \forall \, n \in \mathbb{N}_0 \,,
\end{align}
where higher derivatives are obtained from a matrix recursively defined by
\begin{align}
    Q_0=\mathbf{1}\,, \quad Q_1(x)=-A \,, \quad Q_n(x)=- \partial_x Q_{n-1}(x)+ Q_{n-1}(x)Q_1(x) \,.
\end{align}
Hence we obtain linear inequalities satisfied by the master integral. Imposing them, leads to rigorous bounds on the integrals in the Euclidean region. In favorable cases, one obtains two-sided bounds with rapid convergence.

\subsection{Pedagogical example: The massive bubble integral}
\begin{figure}[h]
\centering 
\includegraphics[width=0.7\textwidth]{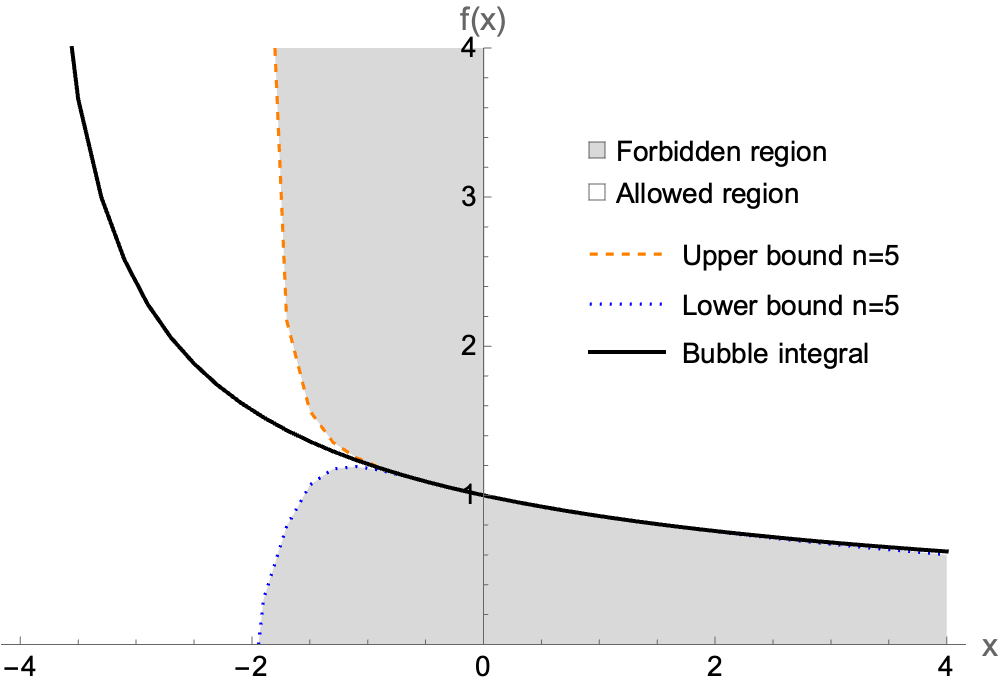}
\caption{Constraints obtained from the CM bootstrap for the massive bubble integral.
The dashed yellow and dashed blue lines correspond to upper and lower bounds, respectively. 
The white regions and gray regions indicate allowed and forbidden regions, respectively.
The solid black line corresponds to the exact function value.}
\label{fig:bubblebootstrap}
\end{figure}

Let us illustrate this on the example of the massive one-loop bubble integral
\begin{align}
f(x) = \int \frac{d^{2}k}{i \pi} 
\frac{1}{
(-k^2+m^2) 
[-(k+p)^2+m^2] 
  } \,.
\end{align}
The corresponding integral family has two master integrals: The bubble and the tadpole integral. For this family, we obtain the differential equation
\begin{align}\label{DEMatrices}
\partial_{x} {\bf f} = A_x \, {\bf f} \,, \quad  {\rm with} \quad
  A_{x} = -\begin{pmatrix}
0 & 0 \\
-\frac{2 }{(4 +x ) x} & \tfrac{2  +x}{(4 +x) x}  
\end{pmatrix}\,,
\end{align}
where we set $m^2=1$ and $-p^2=x$.
Scale invariance allows fixing the value tadpole. Now, one can impose the CM bootstrap to obtain numerical bounds on the value of the bubble integral in the Euclidean region. The results for $n_0=5$ are shown in Fig.~\ref{fig:bubblebootstrap}. There exist three regions with qualitatively different behavior. In the first region, where $x \in (-4,-2)$, we get only a  loose lower bound. For $x \in (-2,0)$, we obtain a two-sided bound, which tightly constrains the value of the function. Note, that the convergence depends strongly on the phase space point. For $x>0$, we get only a lower bound. Interestingly, that bound is very close to the value of the function.

\subsection{Application to multi-loop banana integrals}

We also show, how the CM bootstrap can be applied to banana integrals up to four loops. The resulting bounds are qualitatively similar to those obtained for the bubble integral. Despite the analytic complexity of the function space due to Calabi-Yau-Manifolds, the evaluation time is similar to that of the bubble and only seems to scale with the number of master integrals. This is related to the fact, that the input remains the differential equation matrix, which has purely rational entries. We also compare the evaluation time against \texttt{AMFlow}~\cite{Liu:2022chg}. As for the bubble integral, the convergence of our bootstrap depends strongly on the phase space point. While it is rather slow on one side of the interval with a two-sided bound, it can easily outperform \texttt{AMFlow} on the other side.

\section{Padé approximants for Stieltjes Feynman integrals}
We now show, how to extend these ideas to obtain values in the full cut complex plane using Padé approximants. 
\subsection{Stieltjes functions and Padé approximants}
We first introduce Stieltjes functions, that form a subclass of completely monotonic functions and are defined by admitting an integral representation of the form
\begin{equation} \label{eq:Stieltjesdefinition}
  f(z) = \int_{0}^{\tfrac{1}{R}} \frac{\rho(u)}{1 + u z}\, du \,, 
  \quad 
 {\rm{ with}} \quad \rho(u) \ge 0 \; \forall \; u \in (0, 1/R) \,.
\end{equation}
A key advantage of Stieltjes functions is that their Padé approximants converge rapidly in the complex plane. In particular, $P^{N-1}_{N}(x;x_0)$ and $ P^{N}_{N}(x;x_0)$ provide upper and lower bounds, respectively, to the function for values on the real axis, that converge towards the function value. More generally, they converge towards the function value with rigorous error bounds in the cut complex plane, see~\cite{BGM,Bender}. 

\subsection{Feynman integrals are Stieltjes}
A key result is that certain scalar Feynman integrals satisfy the defining properties of Stieltjes functions.
In Feynman parametrization, Feynman integrals take the schematic form
\begin{align}\label{eq:Feynmanschematic0}
I(\{x_i\}) \sim \int_0^{\infty}  \frac{\prod_i  d\alpha_i~\alpha_i^{\nu_i-1}}{\rm{GL}(1)}  \frac{U^{q}}{F^{e}} \,,
\end{align}
with $U$ being a non-negative homogeneous polynomial in the integration variables $\alpha_i$,
we showed that these integrals are Stieltjes, provided that the exponent of the second Symanzik polynomial satisfies $0<\sum_{i} \nu_i - L D/2\le 1$, where $L$ is the loop order, $D$ is the space-time dimension, and $\nu_i$ are the propagator powers. For any family of Feynman integrals it is usually possible to chose propagator powers that satisfy this constraint. A second condition demands that the Euclidean region exists, i.e. there is a choice of variables for which the second Symanzik polynomial is non-negative, which is always true for sufficiently general kinematics. More precisely, we demand that the second Symanzik polynomial can be put into the form $F =  A \,x+ B$, where $A\ge 0, B\ge 0$. This holds for general planar Feynman integrals, where $x$ can be chosen to be any of the variables $-(p_i + p_{i+1} + \ldots + p_{j-1})^2$ or $m_{i}^2$, with the other variables held fixed in the Euclidean region. In the case of non-planar Feynman integrals, we have provided evidence in the form of examples that the technical condition holds.
For details how these conditions imply the defining properties for Stieltjes functions, see \cite{Ditsch:2025rdx}.

Examples of Feynman integrals that are Stieltjes by this proof are banana integrals in $D=2$, zig-zag integrals in $D=4$ and ladder integrals in $D=6$. 

This observation enables the use of Padé approximants for efficient numerical evaluation. Note that this is valid for all Stieltjes Feynman integrals independently of further analytic properties, such as branch cuts. 
\subsection{Padé approximants as extension of the CM bootstrap: The massive bubble integral}
For a Stieltjes Feynman integral, the CM bootstrap can now be extended to obtain numerical approximations in the full cut complex plane using the following algorithm:
\begin{enumerate}
\item Choose a kinematic point in the Euclidean region where the CM bootstrap provides two-sided bounds with fast convergence.
\item Use the CM bootstrap to compute the value and derivatives of the integral at the starting point.
\item Construct a Taylor expansion around this point.
\item Build the Padé approximants and store them for fast evaluation.
\item Evaluate the approximants across the complex plane.
\end{enumerate}
This method allows for a fast numerical approximation throughout the cut complex plane. The reason is that once the Padé approximants are computed, one only has to evaluate the rational function in different phase space points. It also allows for fast analytic continuation.

\begin{figure}[h]
    \centering
    \includegraphics[width=0.85\linewidth]{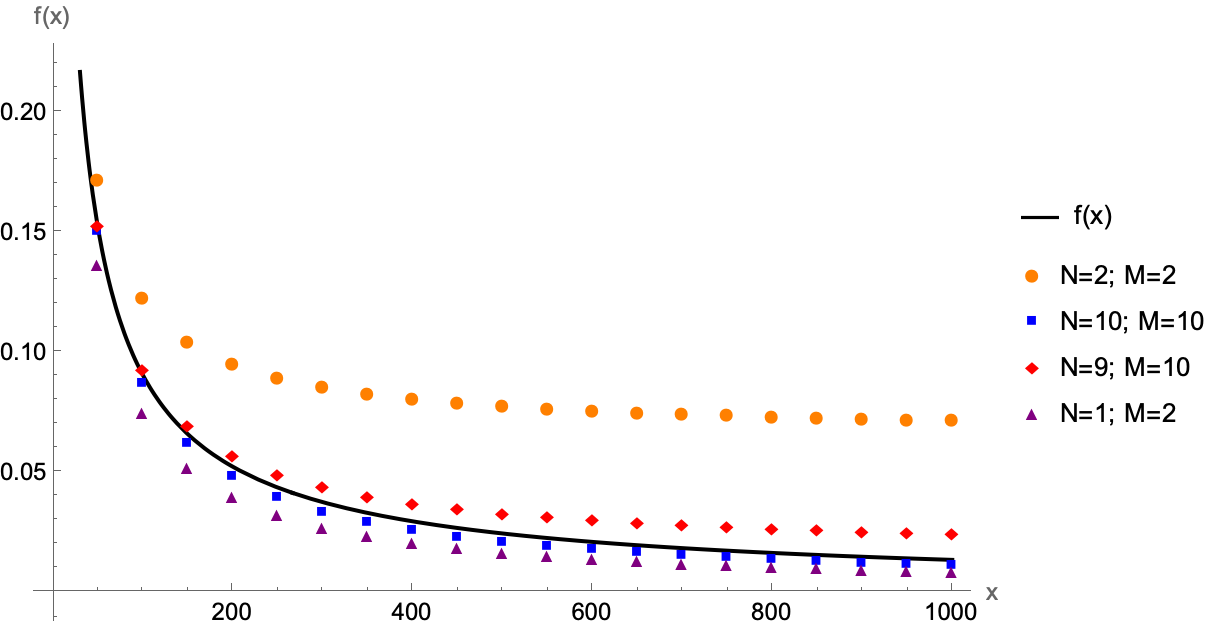}
    \caption{The Padé approximations from the expansion around $x_0= -1/10$  for different values of $N$ and $M$ compared to the massive bubble integral for positive real $x$.
 }
    \label{fig:1a}
\end{figure}

In Fig.~\ref{fig:1a} we show the results of this approach for the bubble integral. Here we chose $x_0=-1/10$ as expansion point, which according to Fig. \ref{fig:bubblebootstrap} is a point with two-sided bounds of good convergence. One can see that the Padé approximants indeed provide a good approximation to the function value for real positive values of $x$ even far away from the expansion point. 

\subsection{Padé approximants beyond differential equations: The 20-loop banana integral}

\begin{figure}[h]
    \centering
   \includegraphics[width=0.6\linewidth]{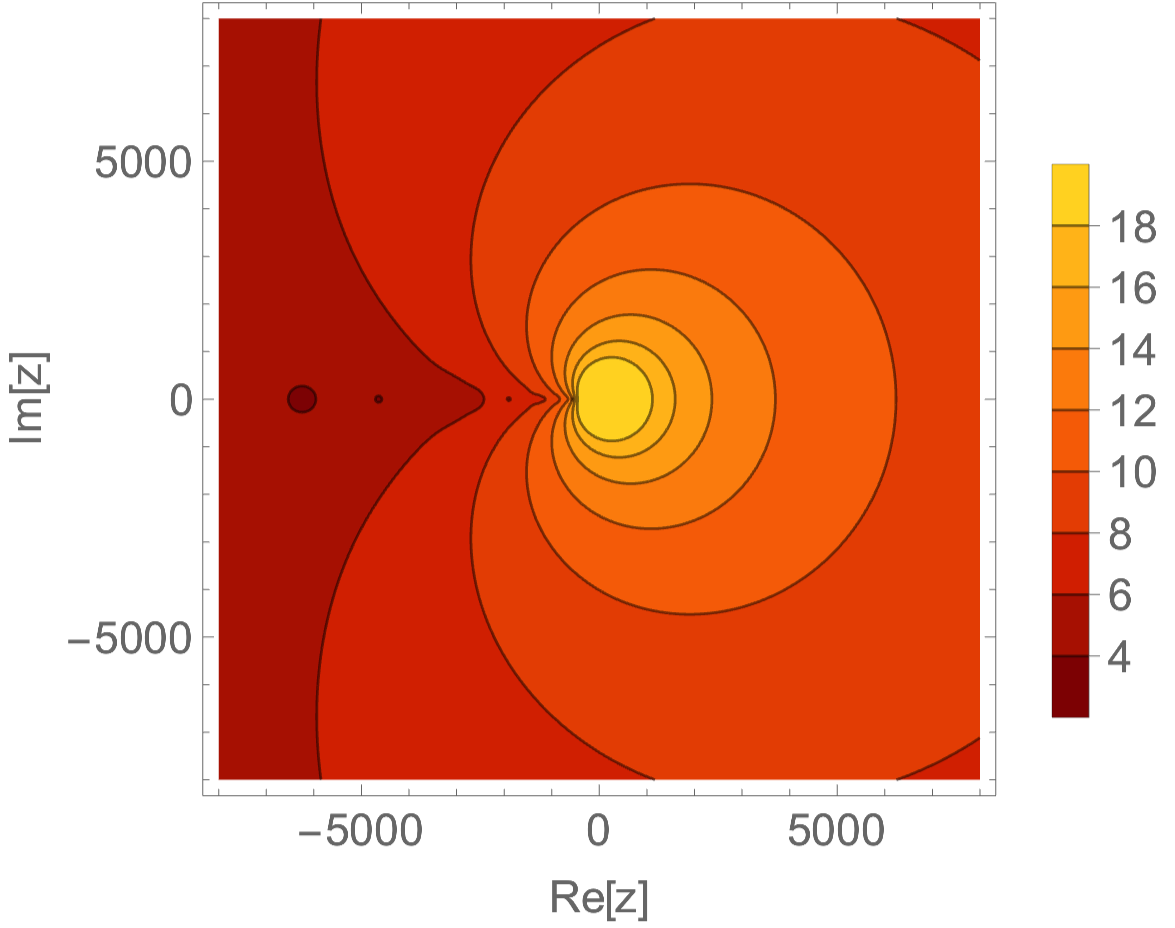}
    \caption{Precision, i.e. digits obtained, of the Padé approximant for $N=10$ of the 20-loop `banana' integral. The light yellow area indicates an agreement of at least 18 digits. The dark red area indicates an agreement of at most four digits.}\label{fig:20_L_banana}
\end{figure}

Our method can also be used to approximate Stieltjes Feynman integrals independent of their differential equation. As long as the Taylor expansion about a convenient point can be obtained in any way, one can use Padé approximants for an efficient approximation. We demonstrate this on the example of the 20-loop banana integral, where we obtain the Taylor expansion about the origin from its Bessel integral representation, see \cite{Groote:2005ay,Pogel:2022vat}. Then we compute the Padé approximants within seconds and evaluate them anywhere in the cut complex plane. Even including only a few derivatives in the Taylor series, this leads to very precise approximations as shown in Fig.~\ref{fig:20_L_banana}. Here we define the precision as the agreement between the $P^{N-1}_{N}(x;x_0)$ and $ P^{N}_{N}(x;x_0)$ Padé approximants.

\section{Conclusion and Outlook}
We presented a novel framework for the numerical evaluation of Feynman integrals based on complete monotonicity and Stieltjes properties. The CM bootstrap yields rigorous bounds without requiring boundary data for values in certain parts of the Euclidean region. We extended this method by showing that Feynman integrals with an appropriate choice of propagator powers are Stieltjes functions. This motivated the use of Padé approximants, which provide efficient numerical approximations with rigorous error bounds.

A natural next step is the extension of this method to integrals in dimensional regularization. This can be achieved by sampling the integrals numerically at different values of the dimensional regulator and reconstructing the associated Laurent expansion. While the Padé approximants can in principle be analytically continued to any point in the cut complex plane, the precision of the expansion quickly decreases close to the branch cut. This makes the continuation to the physical region numerically challenging. Improving the convergence properties in this regime is left to future work. While the present approach already applies to multi-scale integrals by fixing all but one variable, developing a genuinely multivariate extension of the Stieltjes property is also of interest. 

Our method is particularly well suited for integrals in particle physics with many scales, such as electroweak corrections, where IBP-based approaches reach their limits. 
More broadly, complete monotonicity is a structural property shared by a wide class of quantities in quantum field theory. This strongly suggests that the approach developed here extends beyond Feynman integrals and provides a new set of tools for precision calculations in areas such as gravitational physics and cosmology.

\acknowledgments

Funded by the European Union (ERC, UNIVERSE PLUS, 101118787). S.D. was supported by the ERC
Starting Grant 949279 HighPHun. Views and opinions expressed are however those of the authors only and do not necessarily reflect those of the European Union or the European Research Council Executive Agency. Neither the European Union nor the granting authority can be held responsible for them.

\bibliographystyle{JHEP}
\bibliography{ff.bib}

\end{document}